\documentclass[11pt]{article}

\usepackage[a4paper, margin=1.1in]{geometry}
\usepackage{setspace}
\usepackage[T1]{fontenc}
\usepackage[utf8]{inputenc}
\usepackage{mathptmx}    % Times Roman text & math
\usepackage[scaled=0.92]{helvet}
\usepackage{microtype}

\usepackage{amsmath, amssymb, amsthm}

\usepackage{booktabs}
\usepackage{array}
\usepackage{tabularx}
\usepackage{multirow}

\usepackage{enumitem}
\setlist[itemize]{leftmargin=*, topsep=2pt, itemsep=1pt}
\setlist[enumerate]{leftmargin=*, topsep=2pt, itemsep=1pt}

\usepackage{graphicx}
\usepackage{tikz}
\usetikzlibrary{arrows.meta, positioning, calc, shapes, backgrounds, fit}

\usepackage[colorlinks=true, allcolors=black, citecolor=black, urlcolor=black]{hyperref}

\usepackage[font=small, labelfont=bf]{caption}

\usepackage{titlesec}
\titleformat{\section}{\large\bfseries}{}{0pt}{}
\titleformat{\subsection}{\normalsize\bfseries}{}{0pt}{}
\titleformat{\subsubsection}[runin]{\normalsize\bfseries}{}{0pt}{}[.]

\usepackage[numbers, sort&compress]{natbib}
\title{
  \vspace{-0.6cm}
  \LARGE\bfseries
  AI Harness Engineering: A Runtime Substrate for Foundation-Model Software Agents
}

\author{
  \large
  Hailin Zhong$^{1}$, Shengxin Zhu$^{2,\ast}$ \\
  $^{1}$Hong Kong Baptist University, Hong Kong, China\\
  $^{2}$Beijin Normal University, Zhuhai, China\\
  $^{\ast}$Correspondence: \texttt{shengxin.zhu@bnu.edu.cn}
}

\date{}

\begin{document}
% =====================================================================

\maketitle

% ---------------------------------------------------------------------
% Abstract
% ---------------------------------------------------------------------
\begin{abstract}
\noindent
Foundation models have transformed automated code generation, yet autonomous software-engineering agents remain unreliable in realistic development settings. The dominant explanation locates this gap in model capability. We propose a different locus: software-engineering capability emerges from a \emph{model--harness--environment system}, in which a runtime substrate---the harness---mediates how a foundation-model agent observes a project, acts on it, receives feedback, and establishes that a change is complete. We formalize this substrate as an \textbf{AI Harness Engineering} and identify eleven component responsibilities: task specification, context selection, tool access, project memory, task state, observability, failure attribution, verification, permissions, entropy auditing, and intervention recording. We operationalize the harness through a four-level ladder (H0--H3) that progressively exposes runtime support to the agent, and we propose a trace-based evaluation protocol that converts each agent run into an auditable episode package. Applied to a controlled validation task, the framework yields episode packages whose evidence structure varies systematically with harness level: lower levels produce only a final patch, higher levels produce reproduction logs, failure attributions, deterministic requirement checks, and structured verification reports. The framework reframes the central question of autonomous software engineering from whether a foundation model can produce a patch to whether the model--harness--environment system can produce a \emph{verifiably} correct, attributed, and maintainable change. We outline a research program for the runtime systems that foundation-model software agents will require.
\end{abstract}


\vspace{0.4cm}
\hrule
\vspace{0.4cm}

% =====================================================================
% INTRODUCTION
% =====================================================================
\section*{Introduction}

Foundation models have rapidly become capable programming
assistants~\cite{chen2021codex, brown2020gpt3}. They generate functions,
modify files, explain code, write tests, invoke tools, and interact with
software repositories. This progress has energized an ambition that
predates the current generation of models but has only recently become
plausible: autonomous software-engineering agents that take a
high-level development task and carry it through implementation,
testing, verification, and maintenance with minimal human supervision.

Recent benchmarks and agent systems
\cite{jimenez2024swebench, yang2024sweagent, wang2024openhands, xia2024agentless,
       zhang2024autocoderover}
show that the gap between local code generation and complete software
work is real and persistent. A model that writes a correct local patch
may still fail to complete a task. It may inspect the wrong files, apply
a surface-level patch to a user interface while the underlying API
behavior remains broken, run the wrong tests, misinterpret a failure,
forget task state, leave behind obsolete artifacts, or declare success
without sufficient verification. Humans remain involved not primarily
because they are writing every line of code, but because they provide
the missing runtime support: they identify relevant context, explain
repository structure, select tools, interpret feedback, enforce
architectural boundaries, verify behavior, and clean up residue.

The dominant framing of this gap locates it in model capability. On
that view, an agent that fails on a software task does so because the
model is not yet capable enough at coding, reasoning, planning, or
tool use, and the field's task is to train better models or compose
them into more sophisticated agent loops
\cite{shinn2023reflexion, wei2022cot, yao2023react}. We do not contest
that model capability matters. We argue that this framing is
incomplete.

Software engineering is a long-horizon, stateful, tool-mediated,
feedback-driven activity. It depends on context management, project
memory, tool interfaces, execution traces, validation signals,
permissions, rollback, and maintenance discipline. When a foundation
model is placed inside a development environment designed primarily
for human developers, many of these supports remain implicit,
inaccessible, or unstable for the agent. The components that human
developers acquire through socialization, documentation, and
experience are not freely available to a model invocation; they must
be exposed, structured, and traced. \emph{Where they are not, the agent
either improvises or asks the human to fill the gap.}

We therefore propose a different framing. Autonomous software-engineering
capability is an emergent property of a \emph{model--harness--environment
system}, not of the model alone:
\[
  C_{\text{system}} \;=\; F\!\bigl(C_{\text{model}},\, C_{\text{harness}},\,
                                  C_{\text{environment}},\, T\bigr),
\]
where \(C_{\text{model}}\) is the foundation model's latent capability,
\(C_{\text{environment}}\) is what the software environment exposes,
\(C_{\text{harness}}\) is the runtime substrate that mediates between
them, and \(T\) is the task distribution. We call this substrate an
\textbf{AI Harness Engineering}: a runtime layer that surrounds a
foundation-model software agent and manages context, tools, project
memory, task state, observability, failure attribution, verification,
permissions, and maintenance state. The harness is what determines
whether latent model capability becomes \emph{auditable}
software-engineering behavior.

\smallskip
\noindent\textbf{Contributions.} This paper makes four contributions.
\textbf{(i)} We define the AI Harness Engineering as a new research
object distinct from agent--computer interfaces
\cite{yang2024sweagent}, agent frameworks \cite{wu2024autogen,
anthropic2024mcp}, and agent operating systems \cite{mei2024aios}, and
we identify its eleven component responsibilities and five design
principles. \textbf{(ii)} We propose the \emph{H0--H3 harness ladder},
a controlled-visibility ablation that exposes progressively more
runtime support to the agent and makes the harness's contribution
empirically separable from the model's. \textbf{(iii)} We define a
\emph{trace-based evaluation protocol} that records eight classes of
execution evidence---action, tool, context, verification, failure
attribution, intervention, entropy, and outcome---and adjudicates each
agent run by verification autonomy rather than task success alone.
\textbf{(iv)} We instantiate the framework on a controlled validation
task and show that the resulting episode packages differ systematically
in evidence structure across harness levels, with the highest level
producing reproduction logs, failure attributions, requirement-level
verification, and structured verification reports that lower levels do
not.

\smallskip
\noindent\textbf{Why now.} Industrial development practice has
independently begun to converge on harness-like structures around
coding agents. Reports from OpenAI on Codex and from Microsoft on
agent harnesses describe context management, repository knowledge,
observability, tool interfaces, feedback loops, and human attention
as first-class concerns of agent
deployment~\cite{openai2026codex, microsoft2026harness}. These
practitioner accounts confirm that something like a harness exists
and matters. They do not, however, treat the harness as a research
object. They do not define its components, expose its support as a
controlled ablation, or specify the evidence that an episode should
produce. This paper does.

% =====================================================================
% A RUNTIME VIEW OF AUTONOMOUS SOFTWARE ENGINEERING
% =====================================================================
\section*{A runtime view of autonomous software engineering}

\noindent\textbf{From coding ability to software-engineering capability.}
A foundation model that generates correct code fragments, explains an
existing function, or proposes a patch for a localized bug exhibits
\emph{coding ability}. \emph{Software-engineering capability} is more
than this. It is a stateful process involving repository navigation,
context selection, tool use, test execution, failure interpretation,
verification, documentation, and maintenance. The two are commonly
conflated: a model that performs well on isolated code-generation
benchmarks is taken to be a competent software-engineering agent. The
performance gap between local benchmarks and realistic software tasks
\cite{jimenez2024swebench, liu2024agentbench} indicates otherwise.

\begin{figure}[t]
  \centering
  % Figure 1: Model-Harness-Environment System
\begin{tikzpicture}[
  every node/.style={font=\small},
  box/.style={
    draw,
    rounded corners=2pt,
    minimum height=2.6cm,
    minimum width=3.6cm,
    align=center,
    line width=0.7pt
  },
  arrow/.style={
    -{Latex[length=2.2mm]},
    line width=0.7pt
  },
  label/.style={
    font=\scriptsize\itshape,
    align=center
  }
]
  % Foundation Model
  \node[box, fill=gray!5] (model) at (0, 0)
    {\textbf{Foundation Model}\\[2pt]
     \scriptsize latent capability:\\
     \scriptsize code, reasoning, planning,\\
     \scriptsize instruction following};

  % AI Development Harness
  \node[box, fill=gray!15, minimum width=4.6cm, minimum height=4.2cm] (harness) at (6, 0)
    {\textbf{AI Development Harness}\\[3pt]
     \scriptsize task interface\\
     \scriptsize context manager\\
     \scriptsize tool registry\\
     \scriptsize project memory\\
     \scriptsize task state\\
     \scriptsize observability layer\\
     \scriptsize failure attribution\\
     \scriptsize verification protocol\\
     \scriptsize permission boundary\\
     \scriptsize entropy auditor\\
     \scriptsize intervention log};

  % Environment
  \node[box, fill=gray!5] (env) at (12, 0)
    {\textbf{Software Environment}\\[2pt]
     \scriptsize repository, tests,\\
     \scriptsize tools, logs, build,\\
     \scriptsize dependencies};

  % Arrows: model <-> harness
  \draw[arrow] ([yshift=8pt]model.east) -- node[above, label]{prompts,\\context} ([yshift=8pt]harness.west);
  \draw[arrow] ([yshift=-8pt]harness.west) -- node[below, label]{actions,\\tool calls} ([yshift=-8pt]model.east);

  % Arrows: harness <-> environment
  \draw[arrow] ([yshift=8pt]harness.east) -- node[above, label]{tool calls,\\test commands} ([yshift=8pt]env.west);
  \draw[arrow] ([yshift=-8pt]env.west) -- node[below, label]{outputs,\\feedback} ([yshift=-8pt]harness.east);

  % Outcome arrow downward from harness
  \node[below=0.6cm of harness, align=center, font=\scriptsize] (out)
    {\textbf{episode package}\\
     traces, verification report,\\
     failure attribution, entropy audit};
  \draw[arrow] (harness.south) -- (out.north);
\end{tikzpicture}
  \caption{\textbf{The model--harness--environment system.} The
  foundation model provides latent reasoning and coding capability.
  The software environment provides repositories, tests, tools, logs,
  and build affordances. The AI Harness Engineering sits between them,
  mediating context, actions, feedback, and verification evidence.
  Autonomous software-engineering capability is a property of the
  composed system, not of the model alone.}
  \label{fig:system}
\end{figure}
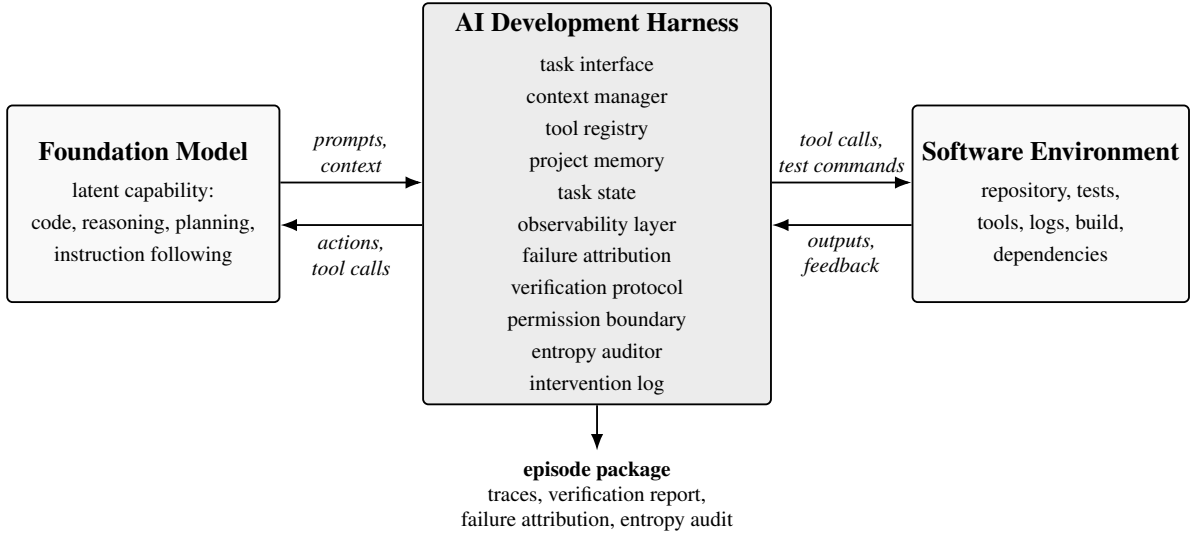

\noindent\textbf{The unit of analysis.} Treating the model alone as the
unit of analysis produces a characteristic attribution error. A failed
autonomous episode is read as a model failure; a successful episode is
credited to the model. But the model rarely acts alone in a realistic
setting. It receives a representation of the task, observes a subset of
the repository, invokes tools through an interface, receives feedback
from tests or commands, and decides when the task is complete. Every
one of these steps is mediated by runtime structure. When that
structure is present and well-designed, the system performs as if the
model is competent. When it is absent or unstable, the same model
appears incompetent. Figure~\ref{fig:system} depicts the unit of
analysis we adopt: the model--harness--environment system, with the
harness as the mediating substrate.

\noindent\textbf{The autonomy gap.} We define the \emph{autonomy gap}
as the difference between a model's apparent local coding ability and
the complete system's ability to perform a software task without
runtime-substituting human help. The gap is not a single failure mode
but a family. An agent may write correct logic but inspect the wrong
files. It may select a relevant test but misinterpret its output. It
may apply a patch to the wrong layer of the architecture. It may
verify the change it made but fail to check that prior behavior is
preserved. It may declare completion without recording evidence. Each
of these is recognizable to practitioners and each maps, in our
framework, to a missing harness responsibility.

\noindent\textbf{Human intervention as runtime signal.} A consequence
of treating the system as the unit of analysis is that human
intervention takes on a new role. In conventional agent evaluation,
human help during an episode is either disallowed (to preserve
autonomy) or treated as noise. We treat it as a diagnostic signal.
When a human tells the agent which file to inspect, this indicates a
missing or inadequate context manager. When a human interprets a test
failure for the agent, this indicates missing observability or
failure-attribution support. When a human verifies the final behavior,
this indicates a missing verification protocol. When a human removes
generated residue, this indicates a missing entropy auditor. We call
such an intervention a \emph{missing-harness human intervention} and
define the missing-harness human intervention rate (M-HIR) accordingly:
\[
  \text{M-HIR} \;=\; \frac{\text{missing-harness interventions}}{
                           \text{total episodes}}.
\]
A harness that lowers M-HIR is one that supplies runtime support the
human would otherwise have to provide.

\noindent\textbf{Failure taxonomy.} The diagnostic value of treating
the system as the unit of analysis depends on being able to
distinguish failure types. We use eight: $F_{\text{context}}$ (the
agent lacks or misuses relevant context); $F_{\text{tool}}$ (a tool
is missing, unstable, or misused); $F_{\text{feedback}}$ (feedback is
unavailable or not interpretable); $F_{\text{verify}}$ (the agent
cannot prove the task requirements are satisfied); $F_{\text{recovery}}$
(the agent cannot recover from a failure); $F_{\text{entropy}}$ (the
agent introduces maintenance burden); $F_{\text{model}}$ (model
reasoning or coding failure despite adequate harness and environment);
and $F_{\text{unknown}}$ (a failure that cannot be confidently
attributed). The taxonomy enables a question that pass/fail evaluation
cannot answer: when an agent fails, \emph{what kind of runtime support
was missing?}

% =====================================================================
% THE AI Harness Engineering
% =====================================================================
\section*{The AI Harness Engineering}

\noindent\textbf{Definition.} An \emph{AI Harness Engineering} is a
runtime substrate surrounding a foundation-model software agent that
manages context, tools, project memory, task state, observability,
failure attribution, verification, permissions, and maintenance state,
so that latent model coding capability becomes auditable
software-engineering behavior.

Four implications follow. First, a harness is external to the model;
it influences model behavior but is not itself the model. Second, a
harness is task-runtime infrastructure: it governs how an agent
\emph{observes} a project, \emph{acts} on it, receives \emph{feedback},
and establishes \emph{completion}. Third, a harness is evaluable: its
components can be exposed, hidden, ablated, traced, and compared.
Fourth, a harness produces \emph{evidence}: why files were chosen,
what tools were used, how failures were attributed, which requirements
were verified, whether human intervention was needed, and what
maintenance burden was introduced.

\noindent\textbf{Five design principles.} A harness should satisfy
five principles:
(P1)~\emph{Explicit runtime resources.} Critical resources---context,
tool affordances, project memory, verification evidence, human
attention, permission boundaries, maintenance state---are exposed and
named rather than left implicit.
(P2)~\emph{Traceable mediation.} The harness records how the agent
selects context, invokes tools, attempts verification, recovers from
failure, and incurs intervention.
(P3)~\emph{Requirement-level verification.} Task completion is bound
to evidence---deterministic checks, targeted tests, regression
attempts, lint, patch review---rather than to a natural-language
assertion.
(P4)~\emph{Attribution before recovery.} A failed observation
produces a classified diagnosis before the agent edits again.
(P5)~\emph{Maintenance and entropy awareness.} The harness records
whether the agent introduced maintenance burden---stale documentation,
dependency churn, generated residue, test weakening, or boundary
violations---rather than treating these as outside the loop.

\noindent\textbf{Eleven component responsibilities.}
Table~\ref{tab:components} enumerates the component responsibilities of
a development harness, the runtime contract each fulfills, the
characteristic failure mode that arises when the responsibility is
absent, and the evidence artifact each produces in a recorded episode.
These eleven are not framework-internal abstractions; they correspond
to identifiable runtime decisions that any agent-on-repository system
must make implicitly, whether or not the harness makes them explicit.

\begin{table}[t]
\centering\small
\caption{\textbf{The eleven component responsibilities of an AI
Development Harness.} For each component we list its runtime contract,
the failure mode that arises when the responsibility is unmanaged, and
the evidence artifact it produces.}
\label{tab:components}
\begin{tabularx}{\linewidth}{@{}lXXl@{}}
\toprule
\textbf{Component} & \textbf{Runtime contract} &
\textbf{Failure when absent} & \textbf{Evidence} \\
\midrule
Task interface & Present objective, requirements, constraints, success criteria & Underspecified goal; wrong-target work & Task record \\
Context manager & Select and expose task-relevant project content & Wrong-file inspection; missed constraints & Context trace \\
Tool registry & Declare available tools and allowed commands & Failed call; unsafe command; repeated timeout & Tool trace \\
Project memory & Provide agent-readable architecture, testing, known-failure knowledge & Repeated rediscovery; wrong-layer fix & Memory references \\
Task state & Maintain hypothesis, inspected files, open questions, next steps & Drift; repeated work; incoherence & Task-state file \\
Observability layer & Expose logs, traces, outputs, runtime errors & Unverifiable success; un-diagnosable failure & Observation log \\
Failure attribution & Separate observation, expected behavior, diagnosis & Random patching after failure & Attribution log \\
Verification protocol & Map task requirements to deterministic evidence & Unverified success; false confidence & Verification trace \\
Permission boundary & Restrict risky actions; expose approval gates & Unsafe invalid episodes & Permission record \\
Entropy auditor & Detect maintenance burden introduced by the agent & Stale docs; dependency churn; residue & Entropy audit \\
Intervention logger & Record human assistance and its avoidability & Invisible human scaffolding & Intervention log \\
\bottomrule
\end{tabularx}
\end{table}

\noindent\textbf{A resource-management view.} Traditional operating
systems manage CPU, memory, files, processes, and devices. A
development harness manages an analogous but distinct set of runtime
resources, summarized in Table~\ref{tab:resources}: context budget,
tool budget, verification evidence, project memory, task state, human
attention, permission boundary, failure signal, entropy budget, and
test-time compute. The analogy serves a single purpose: it identifies
\emph{what must be managed for an agent's behavior to be coherent,
verifiable, and maintainable}. We do not propose an operating system
for AI agents and do not claim that conventional operating system
mechanisms transfer to this setting. The value of the analogy is
strictly that of a resource-management lens.

\begin{table}[t]
\centering\small
\caption{\textbf{Runtime resources managed by an AI Development
Harness.} The harness mediates a set of resources analogous to but
distinct from those managed by conventional operating systems.}
\label{tab:resources}
\begin{tabularx}{\linewidth}{@{}lXX@{}}
\toprule
\textbf{Resource} & \textbf{What it represents} &
\textbf{Failure when unmanaged} \\
\midrule
Context budget & What the agent can see and reason over & Wrong-file selection; missed constraints \\
Tool budget & Which actions the agent can take, when & Inability to inspect, test, or modify \\
Verification evidence & Proof that requirements are satisfied & Premature success claims \\
Project memory & Stable, agent-readable project knowledge & Repeated rediscovery; wrong-layer fixes \\
Task state & Current plan, inspected files, open questions & Drift; incoherent execution \\
Human attention & Cost of human assistance during episode & High missing-harness intervention rate \\
Permission boundary & Allowed and forbidden actions & Unsafe edits; destructive commands \\
Failure signal & Structured feedback from tests, logs, runtime & Random patching; poor recovery \\
Entropy budget & Maintenance burden introduced by the agent & Long-term degradation \\
Test-time compute & Compute spent on verification and exploration & Runaway commands; expensive loops \\
\bottomrule
\end{tabularx}
\end{table}

\noindent\textbf{Positioning.} A harness is distinct from each of the
research objects to which it is most often compared. It is not a
\emph{prompt}: a prompt shapes a single model invocation, while a
harness governs an entire episode. It is not an \emph{agent
framework}~\cite{wu2024autogen, anthropic2024mcp}: an agent framework
provides infrastructure for composing agents and tools, while a
harness is the runtime configuration of supports exposed to a
software agent. It is not an \emph{agent--computer interface}
\cite{yang2024sweagent}: an ACI specifies how an agent acts through
tools, and is one component of a harness. It is not an \emph{agent
operating system} \cite{mei2024aios}: an agent OS targets general agent
scheduling and resource management, while a harness targets a
software-engineering-specific substrate. It is not an \emph{evaluation
harness}: an evaluation harness measures behavior, while a development
harness shapes behavior. And it is not \emph{DevOps or platform
engineering}~\cite{humble2010cd, kim2016devops}: those provide
infrastructure for human and machine development workflows; a
development harness focuses specifically on the runtime interface
between foundation-model agents and software-development environments.
The harness can be built using prompts, agent frameworks, ACIs,
DevOps tools, and operating-system services. The research object is
the configuration of runtime supports exposed to the agent and the
evidence produced during execution.

% =====================================================================
% A CONTROLLED HARNESS LADDER
% =====================================================================
\section*{A controlled harness ladder}

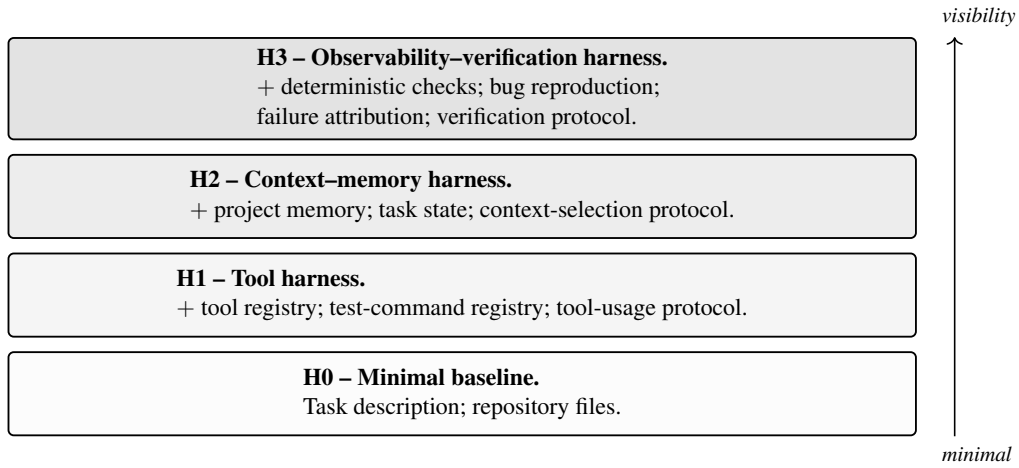
\begin{figure}[t]
  \centering
  % Figure 2: H0-H3 Harness Ladder (clean, with bold headers)
\begin{tikzpicture}[
  every node/.style={font=\footnotesize},
  rung/.style={
    draw,
    rounded corners=2pt,
    minimum width=12cm,
    minimum height=1.1cm,
    align=left,
    inner xsep=10pt,
    inner ysep=4pt,
    line width=0.7pt
  }
]
  % Bottom rung: H0
  \node[rung, fill=gray!2] (h0) at (0, 0)
    {\textbf{H0 -- Minimal baseline.}\\
     Task description; repository files.};

  % H1
  \node[rung, fill=gray!8, above=0.18cm of h0] (h1)
    {\textbf{H1 -- Tool harness.}\\
     $+$ tool registry; test-command registry; tool-usage protocol.};

  % H2
  \node[rung, fill=gray!14, above=0.18cm of h1] (h2)
    {\textbf{H2 -- Context--memory harness.}\\
     $+$ project memory; task state; context-selection protocol.};

  % Top rung: H3
  \node[rung, fill=gray!22, above=0.18cm of h2] (h3)
    {\textbf{H3 -- Observability--verification harness.}\\
     $+$ deterministic checks; bug reproduction;\\
     failure attribution; verification protocol.};

  % Visibility direction labels on the right
  \node[font=\scriptsize\itshape, anchor=south west] at ([xshift=0.2cm]h3.north east)
    {visibility};
  \node[font=\scriptsize\itshape, anchor=north west] at ([xshift=0.2cm]h0.south east)
    {minimal};

  % Single vertical arrow on the right side
  \draw[->, line width=0.6pt] ([xshift=0.5cm]h0.south east)
    -- ([xshift=0.5cm]h3.north east);
\end{tikzpicture}
  \caption{\textbf{The H0--H3 harness ladder.} Each level adds one
  named class of runtime support. Visibility is monotonic: each level
  inherits all artifacts of lower levels. The ladder is a controlled
  ablation that makes the contribution of each runtime-support class
  separable from the others.}
  \label{fig:ladder}
\end{figure}

The harness framework provides a vocabulary; it does not by itself
permit empirical inquiry. To make the harness's contribution
separable from the model's, we need a controlled way to vary runtime
support while holding the task, the repository, and the model fixed.
We propose a four-level ladder, \textbf{H0--H3}, that progressively
exposes runtime support to the agent (Figure~\ref{fig:ladder}).

\noindent\textbf{H0 (Minimal baseline).} The agent receives only the
task description and the repository files. No tool registry, no
project memory, no verification protocol. H0 is the comparison point
against which all other levels are read.

\noindent\textbf{H1 (Tool harness).} H0 plus a tool registry, a
test-command registry, and a tool-usage protocol. H1 makes the action
surface explicit and traceable but does not provide agent-readable
project knowledge or verification discipline.

\noindent\textbf{H2 (Context--memory harness).} H1 plus agent-readable
project memory (architecture, testing conventions, known failures), a
task-state file, and a context-selection protocol. H2 makes context
use explicit and traceable.

\noindent\textbf{H3 (Observability--verification harness).} H2 plus a
deterministic behavioral check registry, a bug-reproduction protocol,
a failure-attribution protocol, a verification protocol, and a
verification report template. H3 makes completion an evidentiary
object rather than an assertion.

\noindent\textbf{Five design requirements.} The ladder satisfies five
requirements. (R1)~\emph{Controlled visibility:} each level exposes
only the artifacts assigned to that level; lower levels do not see
higher-level artifacts. (R2)~\emph{Same task, same repository, same
initial state:} all levels run from the same task and the same
repository state. (R3)~\emph{Traceable runtime support:} when a level
provides a capability, its use is recorded. (R4)~\emph{No hidden
evaluator leakage:} expected files, expected fixes, and evaluator
notes are not visible to the agent at any level. (R5)~\emph{Outcome
comparability:} every level is adjudicated under the same final
outcome taxonomy.

\noindent\textbf{Visibility matrix.} Table~\ref{tab:visibility} states
which artifacts are visible at which level. It is the operational
definition of the ladder.

\begin{table}[t]
\centering\small
\caption{\textbf{Visibility matrix for the H0--H3 ladder.} Each
artifact is either visible (\checkmark) or hidden (---) at a given
level. Visibility is monotonically increasing along the ladder.}
\label{tab:visibility}
\begin{tabular}{@{}lcccc@{}}
\toprule
\textbf{Artifact} & \textbf{H0} & \textbf{H1} & \textbf{H2} & \textbf{H3} \\
\midrule
Task description & \checkmark & \checkmark & \checkmark & \checkmark \\
Repository files & \checkmark & \checkmark & \checkmark & \checkmark \\
\midrule
Tool registry & --- & \checkmark & \checkmark & \checkmark \\
Test-command registry & --- & \checkmark & \checkmark & \checkmark \\
Tool-usage protocol & --- & \checkmark & \checkmark & \checkmark \\
\midrule
\textsc{Agent\_Guide} & --- & --- & \checkmark & \checkmark \\
\textsc{Architecture} & --- & --- & \checkmark & \checkmark \\
\textsc{Testing} guide & --- & --- & \checkmark & \checkmark \\
\textsc{Task\_State} & --- & --- & \checkmark & \checkmark \\
\textsc{Known\_Failures} & --- & --- & \checkmark & \checkmark \\
Context-selection protocol & --- & --- & \checkmark & \checkmark \\
\midrule
Deterministic check registry & --- & --- & --- & \checkmark \\
Bug-reproduction protocol & --- & --- & --- & \checkmark \\
Failure-attribution protocol & --- & --- & --- & \checkmark \\
Verification protocol & --- & --- & --- & \checkmark \\
Verification report template & --- & --- & --- & \checkmark \\
\midrule
Hidden evaluator notes & --- & --- & --- & --- \\
\bottomrule
\end{tabular}
\end{table}

\noindent\textbf{What the ladder measures.} The H0--H3 ladder does not
treat task success as its only outcome. It measures whether the agent
inspected relevant context, used tools, ran tests, reproduced the
failure, attributed it, verified each requirement, preserved prior
behavior, avoided unrelated changes, introduced entropy, and required
human intervention. The next section formalizes these measurements as
a trace-based evaluation protocol.

% =====================================================================
% TRACE-BASED EVALUATION
% =====================================================================
\section*{Trace-based evaluation}

The harness ladder defines what is visible to the agent. The
evaluation protocol defines how each agent run is recorded, verified,
audited, and classified.

\noindent\textbf{Principle.} The principle is simple: autonomous
software-engineering evaluation should measure not only whether a
patch is produced, but whether the model--harness--environment system
produces auditable evidence that the task requirements are satisfied.
A conventional benchmark reports whether a final patch passes tests.
That is useful but insufficient for studying harnesses, because
harnesses shape the \emph{process} by which an agent selects context,
uses tools, interprets failures, verifies behavior, and avoids
maintenance burden. The protocol therefore records both final
outcomes and intermediate evidence.

\noindent\textbf{Episode.} An \emph{episode} is one attempt by a
model--harness--environment system to complete a specified
software-engineering task. An episode is defined by an episode
identifier, the model or agent identity, the harness level, the
repository, the initial commit, the task specification, the visible
artifacts, the allowed tools, the intervention policy, the
verification procedure, and the final outcome rule. The unit of
evaluation is the episode, not a single model response.

\begin{figure}[t]
  \centering
  % Figure 3: Evaluation Pipeline (clean vertical layout, no overlap)
\begin{tikzpicture}[
  every node/.style={font=\footnotesize},
  block/.style={
    draw,
    rounded corners=2pt,
    minimum height=0.9cm,
    minimum width=4.4cm,
    align=center,
    inner sep=4pt,
    line width=0.7pt
  },
  trace/.style={
    draw,
    rounded corners=1.5pt,
    minimum height=0.55cm,
    minimum width=2.6cm,
    align=center,
    inner sep=2pt,
    fill=gray!8,
    line width=0.5pt,
    font=\scriptsize
  },
  arrow/.style={-{Latex[length=2mm]}, line width=0.6pt}
]
  % Top: Input pack
  \node[block, fill=gray!5] (input) at (0, 0)
    {\textbf{Input pack}\\
     \scriptsize task $+$ repository $+$ harness artifacts};

  % Agent episode
  \node[block, fill=gray!18, below=0.5cm of input] (episode)
    {\textbf{Agent episode}\\
     \scriptsize model acts under harness};

  % Eight trace boxes in 4x2 grid (4 cols, 2 rows) below episode
  \node[trace] (t1) at (-4.35, -2.9) {action trace};
  \node[trace, right=0.15cm of t1] (t2) {tool trace};
  \node[trace, right=0.15cm of t2] (t3) {context trace};
  \node[trace, right=0.15cm of t3] (t4) {verification trace};

  \node[trace, below=0.15cm of t1] (t5) {intervention log};
  \node[trace, right=0.15cm of t5] (t6) {failure attribution};
  \node[trace, right=0.15cm of t6] (t7) {entropy audit};
  \node[trace, right=0.15cm of t7] (t8) {outcome record};

  % Patch+report block to the right of traces
  \node[trace, fill=gray!2, below=0.45cm of t6, minimum width=5.5cm] (patch)
    {\textbf{patch} $+$ \textbf{verification report}};

  % Episode package - properly spaced below
  \node[block, fill=gray!18, below=0.45cm of patch] (package)
    {\textbf{Episode package}};

  % Outcome
  \node[block, fill=gray!28, below=0.4cm of package] (outcome)
    {\textbf{Outcome classification}\\
     \scriptsize 5-label taxonomy};

  % Arrows
  \draw[arrow] (input) -- (episode);
  \draw[arrow] (episode.south) -- ++(0, -0.25) -| (t1.north);
  \draw[arrow] (episode.south) -- ++(0, -0.25) -| (t8.north);
  \draw[arrow] (t5.south) |- (patch.west);
  \draw[arrow] (t8.south) |- (patch.east);
  \draw[arrow] (patch) -- (package);
  \draw[arrow] (package) -- (outcome);
\end{tikzpicture}
  \caption{\textbf{The evaluation pipeline.} An input pack
  (task $+$ repository $+$ harness artifacts) is converted through an
  agent episode into an episode package containing eight trace types,
  a patch, and a verification report, which is then classified by a
  five-label final-outcome taxonomy.}
  \label{fig:pipeline}
\end{figure}
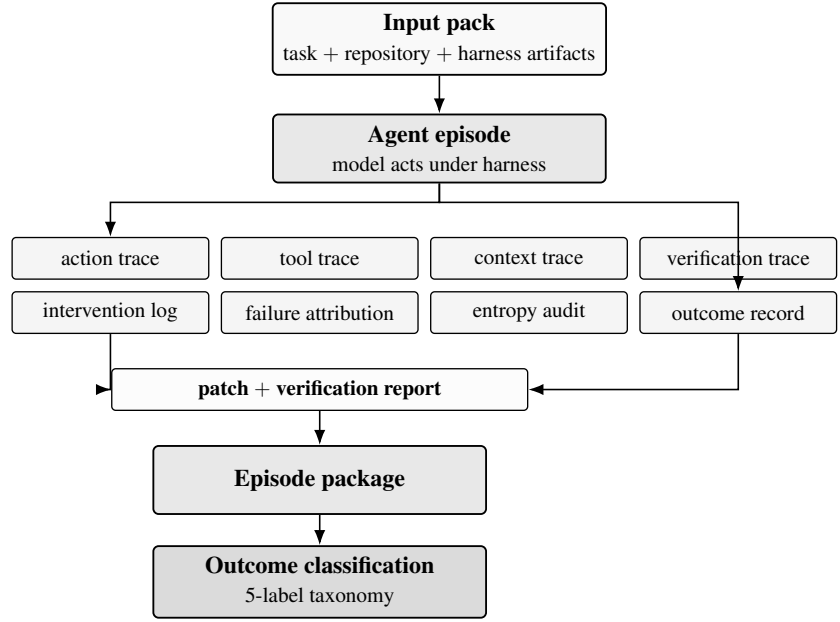

\noindent\textbf{Episode package.} Each episode produces an
\emph{episode package} (Figure~\ref{fig:pipeline}): an auditable
record containing eight trace types plus a patch, a verification
report, a final report, and a final-outcome record.
Table~\ref{tab:traces} maps each trace type to the runtime resource it
captures and the failure type it addresses.

\begin{table}[t]
\centering\small
\caption{\textbf{The eight trace types and what they capture.} Each
trace is associated with one or more runtime resources and the
failure type it makes diagnosable.}
\label{tab:traces}
\begin{tabularx}{\linewidth}{@{}lXX@{}}
\toprule
\textbf{Trace} & \textbf{Runtime resource captured} & \textbf{Failure type addressed} \\
\midrule
action trace & all (sequence of agent operations) & overall episode coherence \\
tool trace & tool budget; test-time compute & $F_{\text{tool}}$ \\
context trace & context budget; project memory & $F_{\text{context}}$ \\
verification trace & verification evidence & $F_{\text{verify}}$ \\
failure-attribution log & failure signal & $F_{\text{verify}}$, $F_{\text{model}}$ \\
intervention log & human attention & all (diagnostic signal) \\
entropy audit & entropy budget & $F_{\text{entropy}}$ \\
outcome record & --- (final classification) & overall adjudication \\
\bottomrule
\end{tabularx}
\end{table}

\noindent\textbf{Trace schemas.} Each trace is line-structured JSON
(JSONL) and admits a compact schema. The action trace records
externally meaningful operations such as \texttt{read\_file},
\texttt{edit\_file}, \texttt{run\_tool}, \texttt{write\_report},
\texttt{update\_task\_state}, \texttt{inspect\_diff}, and
\texttt{declare\_complete}. The tool trace records the command, the
exit code, the duration, the timeout status, the failure type, and
whether the agent recovered. The context trace records which project
memory artifacts were consulted, what they contributed, and whether
the contribution influenced agent decisions. The verification trace
records the type of verification attempted (bug reproduction;
deterministic behavioral check; registered test; targeted test; full
regression; lint; patch review; manual evaluator check), the method,
the result, the requirements covered, and the agent's interpretation.
The failure-attribution log records the observed output, expected
output, failure type, evidence, alternative explanations, and next
diagnostic action. The intervention log records human assistance,
its avoidability, its burden level, and the harness gap it
corresponds to. The entropy audit records categories of
agent-introduced maintenance burden---code, documentation,
dependency, test, file residue, architecture, workflow---together
with a 0--3 severity. The outcome record records final classification
and summary metrics.

\noindent\textbf{Outcome taxonomy.} The final outcome of an episode is
one of five labels. \texttt{autonomous\_\allowbreak verified\_\allowbreak success}:
task requirements are satisfied and sufficient evidence is produced
without missing-harness human intervention.
\texttt{assisted\_\allowbreak verified\_\allowbreak success}:
the final patch is correct, but key progress or verification depended
on human assistance. \texttt{unverified\_\allowbreak success}: the
patch appears correct or the task behavior passes evaluator-side
checks, but the agent did not itself produce evidence sufficient
under the protocol. \texttt{failed}: required behavior fails, tests
fail due to the patch, or no usable patch is produced.
\texttt{unsafe\_\allowbreak invalid}: tests are weakened, unrelated
destructive edits occur, or the task is bypassed. The taxonomy
separates \emph{task behavior} from \emph{evidence quality}: a patch
can be correct but unverified, and a failed patch can be
diagnostically useful.

\noindent\textbf{Deterministic behavioral checks.} The protocol relies
on deterministic behavioral checks to map task requirements to
directly observable outputs. For a validation task, these take the
form of short commands that exercise the corrected behavior, the
preserved valid-input behavior, and the preserved invalid-input
behavior, each with an expected output substring. Deterministic checks
serve two roles: at H3 they are agent-visible harness artifacts that
support the agent's own verification; at all levels they are
evaluator-side adjudication checks that classify the final outcome.
The distinction preserves the ladder while permitting consistent
adjudication.

\noindent\textbf{Metrics.} The protocol enables a family of
process-level metrics: the autonomous verified success rate (AVSR);
the missing-harness human intervention rate (M-HIR); verification
autonomy; context-trace meaningfulness; tool recovery rate; failure
attribution completeness; and entropy delta. These are population-level
quantities that summarize episode packages produced under specified
(model, harness, task, repository) cells.

% =====================================================================
% AN ILLUSTRATIVE CASE
% =====================================================================
\section*{An illustrative case: a controlled validation task}

We illustrate the framework on a controlled task. The task is small by
design: its purpose is to make the ladder and the protocol concretely
inspectable, not to support population-level performance comparisons.
What the case shows is that the H0--H3 ladder is operationally
feasible and that the resulting episode packages differ in evidence
structure in ways that the framework predicts.

\noindent\textbf{Task: repoA-T1.} The repository \texttt{repoA} is a
small login application with a controlled validation defect. The login
flow does not reject an empty password as a validation error; an empty
password instead reaches credential matching and is reported as
\texttt{Invalid credentials}. The task is to modify the application
so that an empty password is rejected with a validation error
containing the substring \texttt{Password is required.}, while
preserving the existing valid-login and invalid-non-empty-credential
behaviors.

\noindent\textbf{Requirements.} The task has five requirements:
(i)~empty password produces a validation error containing
\texttt{"Password is required."};
(ii)~valid credentials (\texttt{alice} / \texttt{correct-password})
still succeed;
(iii)~invalid non-empty credentials still return
\texttt{"Invalid credentials."};
(iv)~a test covers the empty-password behavior;
(v)~existing tests continue to pass, or regression-test instability is
explicitly recorded.

\noindent\textbf{Verification checks.} The evaluator's adjudication
relies on three deterministic behavioral checks: an
empty-password probe, a valid-login probe, and an
invalid-non-empty-credentials probe. Each probe invokes the login
controller directly and is expected to return a specific substring.
Targeted login tests, lint, and full regression are also recorded when
available.

\noindent\textbf{Harness setup.} The same task and the same initial
repository state are evaluated under all four harness levels per the
visibility matrix (Table~\ref{tab:visibility}). Hidden evaluator notes
are not visible to the agent at any level. The agent has no access to
the deterministic check registry below H3.

\smallskip
\noindent\textbf{Outcomes.} Table~\ref{tab:outcomes} summarizes the
result of one execution per harness level. All four levels produce a
working patch; the evidence packages differ in characteristic ways.

\begin{table}[t]
\centering\small
\caption{\textbf{Outcomes and evidence packages across H0--H3 on
repoA-T1.} All four levels execute the task; the distinguishing
characteristic is the evidence structure the episode produces.
``Verified'' here means under the H3 protocol's verification
discipline; lower levels lack a structured verification protocol of
their own.}
\label{tab:outcomes}
\begin{tabularx}{\linewidth}{@{}llX@{}}
\toprule
\textbf{Level} & \textbf{Final outcome} & \textbf{Distinctive evidence produced} \\
\midrule
H0 & \texttt{autonomous\_verified\_success} & patch; evaluator-side deterministic checks pass; full regression succeeds \\
H1 & \texttt{unverified\_success} & patch; tool trace; targeted login test; lint; full regression records a timeout \\
H2 & \texttt{unverified\_success} & H1 evidence plus context trace over project memory; updated task state \\
H3 & \texttt{autonomous\_verified\_success} & H2 evidence plus bug reproduction log; failure attribution log; deterministic requirement checks; structured verification report \\
\bottomrule
\end{tabularx}
\end{table}

\begin{figure}[t]
  \centering
  % Figure 4: H3 Reproduce-Attribute-Fix-Verify-Report Loop (compact)
\begin{tikzpicture}[
  every node/.style={font=\footnotesize},
  hstep/.style={
    draw,
    rounded corners=2pt,
    minimum height=1.1cm,
    minimum width=2.7cm,
    align=center,
    inner sep=3pt,
    line width=0.7pt
  },
  arrow/.style={-{Latex[length=2.2mm]}, line width=0.7pt},
  back/.style={-{Latex[length=2mm]}, line width=0.55pt, dashed, gray}
]
  % Five steps in horizontal flow
  \node[hstep, fill=gray!5] (repro) at (0, 0)
    {\textbf{1. Reproduce}\\[1pt]
     \scriptsize observe failure\\
     \scriptsize vs.\ expected};

  \node[hstep, fill=gray!10, right=0.18cm of repro] (attr)
    {\textbf{2. Attribute}\\[1pt]
     \scriptsize classify failure\\
     \scriptsize type \& evidence};

  \node[hstep, fill=gray!15, right=0.18cm of attr] (fix)
    {\textbf{3. Fix}\\[1pt]
     \scriptsize targeted edit\\
     \scriptsize to attributed layer};

  \node[hstep, fill=gray!20, right=0.18cm of fix] (verify)
    {\textbf{4. Verify}\\[1pt]
     \scriptsize deterministic checks\\
     $+$\scriptsize preserved behavior};

  \node[hstep, fill=gray!25, right=0.18cm of verify] (report)
    {\textbf{5. Report}\\[1pt]
     \scriptsize verification\\
     \scriptsize report $+$ limits};

  % Forward arrows
  \draw[arrow] (repro) -- (attr);
  \draw[arrow] (attr) -- (fix);
  \draw[arrow] (fix) -- (verify);
  \draw[arrow] (verify) -- (report);

  % Back-edge: verify -> attribute (re-attribute if check fails)
  \draw[back] (verify.north) -- ++(0, 0.55)
    -| node[above, font=\scriptsize\itshape, pos=0.25] {re-attribute if verification reveals diagnosis was wrong} (attr.north);
\end{tikzpicture}
  \caption{\textbf{The H3 verification workflow.} H3 binds the agent
  to a five-step discipline: reproduce the failure before editing,
  classify the failure type, apply a targeted fix to the attributed
  layer, verify both required and preserved behavior, and report
  evidence and limitations. A back-edge from verification to
  attribution accommodates the case in which verification reveals
  that the initial diagnosis was wrong.}
  \label{fig:h3loop}
\end{figure}
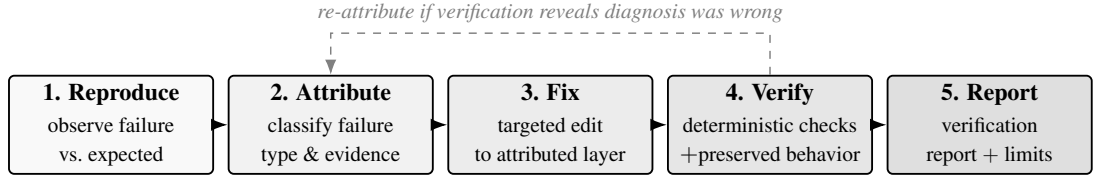

\noindent\textbf{H3 in detail.} H3's distinctive contribution is to
convert task completion into a structured evidentiary object via the
canonical workflow shown in Figure~\ref{fig:h3loop}: reproduce
$\to$ attribute $\to$ fix $\to$ verify $\to$ report. Before editing,
H3 runs the empty-password probe and observes
\begin{center}\small
\verb|{"ok":false,"errors":["Invalid credentials."]}|
\end{center}
against an expected output of
\begin{center}\small
\verb|{"ok":false,"errors":["Password is required."]}|.
\end{center}
The failure is attributed to a validation failure: the empty string
reaches credential matching instead of being rejected by validation.
The fix modifies the validator to reject empty or whitespace-only
passwords, together with a test covering the corrected behavior.
Verification then executes the three deterministic probes plus
targeted tests; a full regression attempt is made and bounded by a
timeout, with the result recorded. The episode concludes with a
verification report linking each requirement to its evidence.

\noindent\textbf{What the case reveals.} Three observations follow
directly from the case. First, \emph{evidence quality varies
systematically with harness level}. Higher levels produce
qualitatively different evidence packages: tool traces appear at H1
and above; context traces appear at H2 and above; reproduction logs,
attribution logs, deterministic-check records, and verification
reports appear only at H3. Second, \emph{tool instability is a
runtime concern, not an incidental nuisance}. The H1--H3 packages
record a full-regression timeout; the protocol surfaces this rather
than hiding it, and H3's verification discipline accommodates it via
deterministic checks bound to specific requirements. Third,
\emph{verification can be a harness responsibility}. Lower levels
treat completion as an assertion; H3 binds the agent to produce
requirement-linked evidence. This is the difference between a patch
and a verifiable change.

% =====================================================================
% DISCUSSION
% =====================================================================
\section*{Implications}

The framework, the ladder, and the protocol together support a
reframing of autonomous software engineering. The central question is
not whether a model can produce a patch but whether the
model--harness--environment system can produce a verifiable, attributed,
maintainable change. Five implications follow.

\noindent\textbf{Verification is a runtime capability.} In many
current workflows, verification is delegated to humans or to external
evaluators: the agent declares completion and someone else checks.
H3 places verification inside the harness. The agent must reproduce
the failure, attribute it, apply a targeted fix, check each
requirement, and report evidence and limitations. This imposes
epistemic discipline on the agent and produces a transferable record
of \emph{why} the change is believed to be correct. A model that
writes code but cannot itself verify behavior remains dependent on
human review; a harness that makes verification explicit reduces that
dependence.

\noindent\textbf{Memory is auditable only when its use is traced.}
Project memory is often treated as a generic benefit for agents.
Providing memory is insufficient; the agent's use of memory must be
inspectable. The context trace records which memory artifact was
consulted, what it contributed, and whether it influenced a decision.
This converts project memory from an invisible prompt ingredient into
an analyzable runtime resource and lets us ask whether the agent
ignored, misunderstood, or correctly applied the memory available to
it.

\noindent\textbf{Failure attribution separates diagnosis from
action.} Agent systems frequently move from a failed observation
directly to a new edit. The result is lucky fixes when the new edit
happens to address the underlying cause, and random patching when it
does not. H3 inserts an attribution step: the agent records observed
output, expected output, inferred failure type, supporting evidence,
alternative explanations, and the next diagnostic action. The
attribution log is auditable: an evaluator can inspect whether the
agent's diagnosis was reasonable before scrutinizing the patch.

\noindent\textbf{Tool stability is a harness problem.} Autonomous
agents act in real environments where commands can hang, tests can be
flaky, and full regression suites can be expensive or unstable.
Human developers adapt: they choose targeted tests, add timeouts,
inspect logs, or report uncertainty. The same adaptation must be
available to the agent. The protocol's distinction between
deterministic behavioral checks, targeted tests, full regression
attempts, and lint---each recorded with its outcome and any
instability---makes tool stability an analyzable runtime resource
rather than an incidental engineering nuisance.

\noindent\textbf{Entropy is part of autonomous engineering.}
Autonomous agents do not only produce solutions. They can also
produce residue: redundant code, stale documentation, unnecessary
dependencies, weakened tests, debug scripts, inconsistent task notes,
or architecture violations. These do not break the immediate task but
degrade the project over time. The entropy auditor places this concern
inside the harness rather than outside the loop. As autonomous agents
take on more sustained software work, entropy management is likely to
become as important as the immediate code change.

\noindent\textbf{Toward AI-native development environments.} Current
software repositories are designed for human developers: they assume
that the reader can infer architecture from convention, remember
testing practices, interpret test failures, and clean up residue.
None of these assumptions hold for a foundation-model agent. The
harness framework suggests that future repositories will need
explicit, agent-readable affordances: architecture maps, testing
guides, deterministic check registries, task-state files, failure
templates, verification report templates, entropy dashboards,
permission manifests, and intervention logs. The design question
shifts from \emph{what should a software repository contain for human
developers} to \emph{what should it expose so that an agent can work
on it reliably}.

% =====================================================================
% OUTLOOK
% =====================================================================
\section*{Outlook}

The framework opens an empirical program rather than concluding one.
The harness ladder is a controlled-ablation instrument; the protocol
produces episode packages that admit population-level analysis. We
identify six directions in which the program naturally extends.

\noindent\emph{Multi-task evaluation.} The harness ladder is designed
to be reusable across task classes. A balanced task suite would stress
different harness components: validation tasks stress the verification
protocol; UI-behavior tasks stress observability; dependency-cleanup
tasks stress the entropy auditor; refactoring tasks stress project
memory and architecture guidance; flaky-test diagnosis stresses
failure attribution and recovery; long-feature implementation
stresses task state and context management; permission-sensitive tasks
stress the permission boundary. Each task class illuminates a
different facet of the harness.

\noindent\emph{Multi-model evaluation.} Harness effects may interact
with model capability. A stronger model may localize files with
minimal guidance; a weaker model may depend heavily on project memory.
Some models may follow structured verification protocols well; others
may not. The
$\text{models} \times \text{harness levels} \times \text{tasks}$
design separates model effects from harness effects and reveals which
harness components are model-agnostic and which are
model-dependent.

\noindent\emph{Quantitative metrics.} The protocol defines AVSR,
M-HIR, verification autonomy, tool recovery rate, failure attribution
completeness, and entropy delta. With a multi-task multi-model design,
these become statistically estimable quantities with confidence
intervals, supporting hypothesis testing of harness contributions.

\noindent\emph{Long-horizon evaluation.} Real software work spans
many episodes on the same repository. Entropy accumulates; task state
becomes more useful or more contradictory; project memory either
ages well or rots. A long-horizon evaluation runs a sequence of tasks
under each harness level, preserves repository state between tasks,
audits entropy after each episode, and measures how prior episodes
affect later success. This is the natural test of whether richer
harnesses reduce long-term maintenance burden.

\noindent\emph{AI-native repository design.} If a harness benefits
from agent-readable affordances, repository design itself becomes a
research question. What documentation structure best supports an
agent's context selection? What test-command registry shape best
supports recovery from instability? What architecture-map format best
prevents wrong-layer fixes? These are questions about repository
artifacts, not about models or agents.

\noindent\emph{Runtime systems for agent-first development.} The
broader implication is that autonomous software engineering will
require runtime systems analogous to operating systems, but
specialized for foundation-model agents: systems that manage context,
tools, memory, verification, permissions, failure recovery, human
oversight, entropy, cost, and risk. The AI Harness Engineering names
the missing layer. The next stage is to build it.

% =====================================================================
% METHODS
% =====================================================================
\section*{Methods}

\noindent\textbf{Repository construction.} The repository
\texttt{repoA} is a small Node.js login application constructed
specifically to support the H0--H3 ladder. It contains an API
controller, a validator, an authentication service, a UI layer, and
an existing test suite. The validation defect is localized to the
validator: it accepts empty-string passwords. The defect is chosen
to be objectively checkable through three deterministic behavioral
probes, to require modifying a specific architectural layer
(rejecting it elsewhere would be a wrong-layer fix), and to admit
verification via a structured workflow.

\noindent\textbf{Harness instantiation.} Each harness level is
instantiated by exposing exactly the artifacts listed in
Table~\ref{tab:visibility}. H0 receives a task description file and
the unmodified repository tree. H1 additionally receives a tool
registry, a test-command registry, and a tool-usage protocol, placed
under a top-level \texttt{harness/} directory. H2 additionally
receives an agent guide, an architecture document, a testing guide, a
task-state file, a known-failures file, and a context-selection
protocol. H3 additionally receives a deterministic check registry, a
bug-reproduction protocol, a failure-attribution protocol, a
verification protocol, and a verification report template. Evaluator
notes, including the expected attribution and the expected fix, are
kept in a separate evaluator pack not visible to the agent at any
level.

\noindent\textbf{Trace recording.} Traces are recorded as JSONL files
following the schemas summarized in the main text. The action trace
captures externally meaningful operations rather than every token of
internal reasoning. The tool trace records every command invocation
with exit code, duration, timeout status, and recovery status. The
context trace records every project-memory artifact consulted by the
agent with a structured \texttt{contribution} field. The verification
trace records every verification attempt with type, method, result,
covered requirements, and the agent's interpretation. The
failure-attribution log is required only at H3. The intervention log
records every human action with an avoidability classification and a
corresponding harness-gap label. The entropy audit is produced at the
end of each episode and classifies any agent-introduced residue.

\noindent\textbf{Deterministic behavioral checks.} The three checks
exercise the login controller directly with three input cases. The
empty-password probe invokes the controller with
\texttt{username = "alice"} and \texttt{password = ""}; the expected
output contains the substring \texttt{"Password is required."} The
valid-login probe uses \texttt{password = "correct-password"} and
expects the substring \texttt{"ok":true}. The
invalid-non-empty-credentials probe uses
\texttt{password = "wrong-password"} and expects the substring
\texttt{"Invalid credentials."} Each probe is executed as a short
Node.js invocation that requires the login controller, calls it with
the test input, and prints the JSON response.
These checks are agent-visible only at H3 (via the deterministic
check registry); they are evaluator-side adjudication checks at all
levels.

\noindent\textbf{Outcome adjudication.} The evaluator applies the
deterministic checks, targeted login tests, lint, and a bounded
full-regression attempt to every episode. The five outcome labels are
assigned by rule: a patch that passes all deterministic checks and is
accompanied by a verification protocol that maps requirements to
evidence is \texttt{autonomous\_verified\_success}; a patch that passes
the deterministic checks without an internal verification protocol is
\texttt{unverified\_success}; an unsuccessful patch is
\texttt{failed}; a patch that weakens tests or introduces unrelated
destructive edits is \texttt{unsafe\_invalid}; a successful patch that
required substantive human assistance is
\texttt{assisted\_verified\_success}. The taxonomy distinguishes
\emph{task behavior} from \emph{evidence quality}.

\noindent\textbf{Full regression handling.} Full regression is
attempted under a strict timeout. If full regression times out or
triggers platform-level instability, the timeout is recorded; full
regression is not silently retried, and a missing full-regression
result is not silently treated as success. At H3, a full-regression
timeout does not by itself prevent
\texttt{autonomous\_verified\_success} when deterministic requirement
coverage is complete and the limitation is reported in the
verification report.

\noindent\textbf{Compute environment.} Each episode is executed in an
isolated workspace with the repository checked out at a fixed initial
commit. Tool invocations and test commands run in subprocesses with
explicit timeouts. The agent's compute environment is identical
across harness levels except for the harness artifacts described
above.

\bibliography{references}

@inproceedings{jimenez2024swebench,
  author    = {Jimenez, Carlos E. and Yang, John and Wettig, Alexander and
               Yao, Shunyu and Pei, Kexin and Press, Ofir and Narasimhan, Karthik},
  title     = {{SWE-bench}: Can Language Models Resolve Real-World {GitHub} Issues?},
  booktitle = {International Conference on Learning Representations (ICLR)},
  year      = {2024}
}

@inproceedings{yang2024sweagent,
  author    = {Yang, John and Jimenez, Carlos E. and Wettig, Alexander and
               Lieret, Kilian and Yao, Shunyu and Narasimhan, Karthik and Press, Ofir},
  title     = {{SWE-agent}: Agent--Computer Interfaces Enable Automated Software Engineering},
  booktitle = {Advances in Neural Information Processing Systems (NeurIPS)},
  year      = {2024}
}

@inproceedings{yao2023react,
  author    = {Yao, Shunyu and Zhao, Jeffrey and Yu, Dian and Du, Nan and
               Shafran, Izhak and Narasimhan, Karthik and Cao, Yuan},
  title     = {{ReAct}: Synergizing Reasoning and Acting in Language Models},
  booktitle = {International Conference on Learning Representations (ICLR)},
  year      = {2023}
}

@inproceedings{liu2024agentbench,
  author    = {Liu, Xiao and Yu, Hao and Zhang, Hanchen and Xu, Yifan and
               Lei, Xuanyu and Lai, Hanyu and Gu, Yu and Ding, Hangliang and
               Men, Kaiwen and Yang, Kejuan and Zhang, Shudan and Deng, Xiang
               and Zeng, Aohan and Du, Zhengxiao and Zhang, Chenhui and Shen,
               Sheng and Zhang, Tianjun and Su, Yu and Sun, Huan and Huang,
               Minlie and Dong, Yuxiao and Tang, Jie},
  title     = {{AgentBench}: Evaluating {LLMs} as Agents},
  booktitle = {International Conference on Learning Representations (ICLR)},
  year      = {2024}
}

@inproceedings{wu2024autogen,
  author    = {Wu, Qingyun and Bansal, Gagan and Zhang, Jieyu and Wu, Yiran and
               Li, Beibin and Zhu, Erkang and Jiang, Li and Zhang, Xiaoyun
               and Zhang, Shaokun and Liu, Jiale and Awadallah, Ahmed Hassan
               and White, Ryen W. and Burger, Doug and Wang, Chi},
  title     = {{AutoGen}: Enabling Next-Gen {LLM} Applications via
               Multi-Agent Conversation},
  booktitle = {COLM},
  year      = {2024}
}

@misc{anthropic2024mcp,
  author       = {{Anthropic}},
  title        = {Introducing the {Model Context Protocol}},
  howpublished = {Anthropic blog},
  year         = {2024}
}

@article{mei2024aios,
  author  = {Mei, Kai and Li, Zelong and Xu, Shuyuan and Ye, Ruosong and
             Ge, Yingqiang and Zhang, Yongfeng},
  title   = {{AIOS}: {LLM} Agent Operating System},
  journal = {arXiv preprint arXiv:2403.16971},
  year    = {2024}
}

@article{chen2021codex,
  author  = {Chen, Mark and Tworek, Jerry and Jun, Heewoo and Yuan, Qiming and
             Pinto, Henrique Ponde de Oliveira and Kaplan, Jared and Edwards, Harri
             and Burda, Yuri and Joseph, Nicholas and Brockman, Greg and others},
  title   = {Evaluating Large Language Models Trained on Code},
  journal = {arXiv preprint arXiv:2107.03374},
  year    = {2021}
}

@misc{openai2026codex,
  author       = {{OpenAI}},
  title        = {{Codex}: Lessons from Building Agent-First Software},
  howpublished = {OpenAI engineering report},
  year         = {2026}
}

@misc{microsoft2026harness,
  author       = {{Microsoft}},
  title        = {Building Agent Harnesses for Developer Tools},
  howpublished = {Microsoft engineering blog},
  year         = {2026}
}

@inproceedings{wang2024openhands,
  author    = {Wang, Xingyao and Li, Bowen and Song, Yufan and Xu, Frank F.
               and Tang, Xiangru and Zhuge, Mingchen and Pan, Jiayi and
               Song, Yueqi and Li, Bowen and Singh, Jaskirat and Tran,
               Hoang H. and Li, Fuqiang and Ma, Ren and Zheng, Mingzhang
               and Qian, Bill and Shao, Yanjun and Muennighoff, Niklas and
               Zhang, Yizhe and Hui, Binyuan and Lin, Junyang and Brennan,
               Robert and Peng, Hao and Ji, Heng and Neubig, Graham},
  title     = {{OpenHands}: An Open Platform for {AI} Software Developers as
               Generalist Agents},
  booktitle = {International Conference on Learning Representations (ICLR)},
  year      = {2025}
}

@article{xia2024agentless,
  author  = {Xia, Chunqiu Steven and Deng, Yinlin and Dunn, Soren and Zhang, Lingming},
  title   = {{Agentless}: Demystifying {LLM}-based Software Engineering Agents},
  journal = {arXiv preprint arXiv:2407.01489},
  year    = {2024}
}

@article{zhang2024autocoderover,
  author  = {Zhang, Yuntong and Ruan, Haifeng and Fan, Zhiyu and Roychoudhury, Abhik},
  title   = {{AutoCodeRover}: Autonomous Program Improvement},
  journal = {arXiv preprint arXiv:2404.05427},
  year    = {2024}
}

@article{shinn2023reflexion,
  author  = {Shinn, Noah and Cassano, Federico and Berman, Edward and
             Gopinath, Ashwin and Narasimhan, Karthik and Yao, Shunyu},
  title   = {{Reflexion}: Language Agents with Verbal Reinforcement Learning},
  journal = {Advances in Neural Information Processing Systems},
  year    = {2023}
}

@inproceedings{wei2022cot,
  author    = {Wei, Jason and Wang, Xuezhi and Schuurmans, Dale and Bosma,
               Maarten and Ichter, Brian and Xia, Fei and Chi, Ed H. and Le,
               Quoc V. and Zhou, Denny},
  title     = {Chain-of-Thought Prompting Elicits Reasoning in Large
               Language Models},
  booktitle = {Advances in Neural Information Processing Systems},
  year      = {2022}
}

@book{humble2010cd,
  author    = {Humble, Jez and Farley, David},
  title     = {Continuous Delivery: Reliable Software Releases through
               Build, Test, and Deployment Automation},
  publisher = {Addison-Wesley},
  year      = {2010}
}

@book{kim2016devops,
  author    = {Kim, Gene and Humble, Jez and Debois, Patrick and Willis, John},
  title     = {The {DevOps} Handbook},
  publisher = {IT Revolution Press},
  year      = {2016}
}

@article{brown2020gpt3,
  author  = {Brown, Tom B. and Mann, Benjamin and Ryder, Nick and Subbiah,
             Melanie and Kaplan, Jared and Dhariwal, Prafulla and others},
  title   = {Language Models are Few-Shot Learners},
  journal = {Advances in Neural Information Processing Systems},
  year    = {2020}
}

\end{document}